\documentclass[aps,showpacs,preprintnumbers,amsmath, amssymb]{revtex4}

\usepackage{float}
\usepackage{graphics,epsfig}
\usepackage{graphicx}
\usepackage{dcolumn}
\usepackage{bm}

\begin{document}
\baselineskip=0.8 cm
\title{{\bf Hawking radiation in a $d$-dimensional static spherically-symmetric black Hole surrounded by quintessence}}
\author{Songbai Chen}
\email{chsb@fudan.edu.cn} \affiliation{ Institute of Physics and
Department of Physics, Hunan Normal University, Changsha, Hunan
410081, P. R. China \\ Key Laboratory of Low Dimensional Quantum
Structures \\ and Quantum Control of Ministry of Education, Hunan
Normal University, Changsha, Hunan 410081, P. R. China\\Department
of Physics, Fudan University, Shanghai 200433, P. R. China}

\author{Bin Wang}
\email{wangb@fudan.edu.cn} \affiliation{Department of Physics,
Fudan University, Shanghai 200433, P. R. China}

\author{ Rukeng Su}
\email{rksu@fudan.ac.cn}
 \affiliation{China Center of Advanced Science and Technology (World Laboratory),
P.B.Box 8730, Beijing 100080, People¡¯s Republic of China
\\ Department of Physics, Fudan University, Shanghai 200433, P. R. China}

\vspace*{0.2cm}
\begin{abstract}
\baselineskip=0.6 cm
\begin{center}
{\bf Abstract}
\end{center}

We present a solution of Einstein equations with quintessential
matter surrounding a $d$-dimensional black hole, whose asymptotic
structures are determined by the state of the quintessential
matter. We examine the thermodynamics of this black hole and find
that the mass of the black hole depends on the equation of state
of the quintessence, while the first law is universal.
Investigating the Hawking radiation in this black hole background,
we observe that the Hawking radiation dominates on the brane in
the low-energy regime. For different asymptotic structures caused
by the equation of state of the quintessential matter surrounding
the black hole, we learn that the influences by the state
parameter of the quintessence on Hawking radiation are different.

\end{abstract}

 \maketitle
\newpage
\section{Introduction}

String theory predicts the existence of extra dimensions. This
inspired a lot of interest to study whether extra dimensions can
be observed, which can present the signature of string and the
correctness of string theory. A great deal of effort has been
expanded for the detection of extra dimensions. One among them is
the study of perturbations around braneworld black holes. It has
been argued that the extra dimension can imprint in the wave
dynamics in the branworld black hole background \cite{1, 2, 3, 4}.
Another chief possibility to observe the extra dimension is the
spectrum of Hawking radiation which is expected to be detected in
particle accelerator experiments \cite{5}-\cite{17}. Recently
through the study of Hawking radiation from squashed Kaluza-Klein
black holes \cite{14,15,16}, it was argued that the luminosity of
Hawking radiation can tell us the size of the extra dimension
which opens a window to observe extra dimensions.

Recent astronomical observations strongly suggest that our universe
is currently undergoing a phase of accelerated expansion, likely
driven by some exotic component called dark energy. Despite the
mounting observational evidence, the nature and origin of dark
energy remains elusive and it has become a source of vivid debate.
Quintessence is one candidate for the dark energy, which has
negative pressure. If the quintessence exists everywhere in the
universe, it can cause the universe to accelerate. Besides with the
quintessence around a black hole, the black hole spacetime will be
deformed. The Einstein equations for the static
spherically-symmetric quintessence surrounding a black hole in four
dimensions were studied in \cite{18}. It was found that the
condition of additivity and linearity in the energy-momentum tensor
allows one to get correct limit to the known solutions for the
electromagnetic static field and for the case of cosmological
constant.

In this work we first extend \cite{18} to the solution of Einstein
equations with quintessential matter surrounding a $d$-dimensional
black hole by assuming that quintessence is not only on the brane
but full in the bulk. We get a new $d$-dimensional black hole, whose
asymptotic structures are determined by the state of the
quintessential matter surrounding the black hole. We examine the
thermodynamics of this black hole and find that the mass of the
black hole depends on the equation of state of the quintessence,
while the first law keeps the same form independent of the
dimensions and the state of quintessence. Investigating Hawking
radiation in this black hole background, we observe that Hawking
radiation dominates on the brane. For different asymptotic
structures caused by the equation of state of the quintessential
matter surrounding the black hole, we learn that the influences by
the state parameter of the quintessence on Hawking radiation are
different. The signature of the dimension in Hawking radiation is
also presented.

\section{$d$-dimensional static spherically-symmetric black holes surrounded by quintessence}

We study the Einstein equation for a static spherically-symmetric
black hole surrounded by quintessence in $d$-dimensions. The
$d$-dimensional static black hole is described by
\begin{eqnarray}
ds^2=e^{\nu}dt^2-e^{\lambda}dr^2-r^2d\theta^2_1-r^2\sin^2{\theta_1}d\theta^2_2-\cdots-r^2\sin^2{\theta_1}
\cdots\sin^2{\theta_{d-3}}d\theta^2_{d-2},\label{1}
\end{eqnarray}
where $\nu$ and $\lambda$ are function of radial coordinate $r$.
The energy-momentum tensor of the quintessence in the static
spherically symmetric state can be written as \cite{18}
\begin{eqnarray}
T^{\;t}_t=A(r), \;\;\;\;\;\;T^{\;j}_t=0,\;\;\;\;\;\;
T^{\;j}_i=C(r)r_ir^j+B(r)\delta^{\;j}_i.
\end{eqnarray}
After averaging over the angles of isotropic state we get
\begin{eqnarray}
\langle T^{\;j}_i\rangle=D(r)\delta^{\;j}_i,\;\;\;\;\;\;\;
D(r)=-\frac{1}{d-1}C(r)r^2+B(r).
\end{eqnarray}
For quintessence, we have
\begin{eqnarray}
D(r)=-\omega_qA(r).
\end{eqnarray}
Thus, in terms of density $A(r)$, we can get the expression of
$D(r)$ for fixed state parameter $\omega_q$. As in Ref.\cite{18},
the appropriate constant coefficient $C(r)/B(r)$ is defined by the
condition of additivity and linearity.

The Einstein equations of metric (\ref{1}) have the form
\begin{eqnarray}
2T^{\;t}_t&=&\frac{d-2}{2}\bigg[-e^{-\lambda}\bigg(\frac{d-3}{r^2}-\frac{\lambda'}{r}\bigg)+\frac{d-3}{r^2}\bigg],\\
2T^{\;r}_r&=&\frac{d-2}{2}\bigg[-e^{-\lambda}\bigg(\frac{d-3}{r^2}+\frac{\nu'}{r}\bigg)+\frac{d-3}{r^2}\bigg],
\\
2T^{\;\theta_1}_{\theta_1}&=&2T^{\;\theta_2}_{\theta_2}=\cdots
2T^{\;\theta_{d-2}}_{\theta_{d-2}}\nonumber\\&=&
-\frac{e^{-\lambda}}{2}\bigg[\nu''+\frac{\nu'^2}{2}-\frac{\lambda'\nu'}{2}
+\frac{(d-3)(\nu'-\lambda')}{r}+\frac{(d-3)(d-4)}{r^2}\bigg]+\frac{(d-3)(d-4)}{2r^2},
\end{eqnarray}
where the prime denotes the derivative with respect to $r$.

The appropriate general expression of the energy-momentum tensor
of quintessence in the $d-$dimensional spherically-symmetric
spacetime is given by
\begin{eqnarray}
T^{\;t}_t=\rho_q(r),\;\;\;\;\;\;\;T^{\;j}_i=\rho_q(r)\alpha
\bigg\{-[1+(d-1)B(r)]\frac{r_ir^j}{r_nr^n}+B\delta^{\;j}_i\bigg\}.\label{ent}
\end{eqnarray}
This leads the spatial part of the energy-momentum tensor in
proportional to the time component with the arbitrary parameter
$B(r)$ which depends on the internal structure of quintessence.
After taking isotropic average over the angles, we obtain
\begin{eqnarray}
\langle r_ir^j\rangle &=&\frac{1}{d-1}r_nr^n\delta^{\;j}_i,\\
\langle
T^{\;j}_i\rangle&=&-\rho_q(r)\frac{\alpha}{d-1}\delta^{\;j}_i=-p_q\delta^{\;j}_i.
\end{eqnarray}
From the state equation $p_q=\omega_q\rho_q$, it is easy to see
\begin{eqnarray}
\omega_q=\frac{\alpha}{d-1}.
\end{eqnarray}
For quintessence, we have $-1<\omega_q<0$ and  $-(d-1)<\alpha<0$.

As in Ref.\cite{18}, we can define a principle of additivity and
linearity by the equality
\begin{eqnarray}
T^{\;t}_t=T^{\;r}_r\Longrightarrow\lambda+\nu=0.
\end{eqnarray}
And then substituting
\begin{eqnarray}
\lambda=-\ln{f},
\end{eqnarray}
we can obtain the linear differential equations in $f$
\begin{eqnarray}
T^{\;t}_t&=&T^{\;r}_r=-\frac{d-2}{4r^2}[rf'+(d-3)(f-1)],\label{t1}\\
T^{\;\theta_1}_{\theta_1}&=&T^{\;\theta_2}_{\theta_2}=\cdots
T^{\;\theta_{d-2}}_{\theta_{d-2}}=
-\frac{1}{4r^2}[r^2f''+2(d-3)rf'+(d-4)(d-3)(f-1)]\label{t2}.
\end{eqnarray}
From equations (\ref{ent}) and (\ref{t1}), we can fix the free
parameter $B$ in the energy-momentum tensor for the matter
\begin{eqnarray}
B=-\frac{(d-1)\omega_q+1}{(d-1)(d-2)\omega_q}.
\end{eqnarray}
Thus the energy-momentum tensor (\ref{ent}) has the form
\begin{eqnarray}
T^{\;t}_t&=&T^{\;r}_r=\rho_q,\label{t3} \\
T^{\;\theta_1}_{\theta_1}&=&T^{\;\theta_2}_{\theta_2}=\cdots
T^{\;\theta_{d-2}}_{\theta_{d-2}}=-\frac{1}{d-2}\rho_q[(d-1)\omega_q+1]\label{t4}.
\end{eqnarray}
Making use of equations (\ref{t1})-(\ref{t2}) and
(\ref{t3})-(\ref{t4}), we obtain a differential equation for $f$
\begin{eqnarray}
r^2f''+[(d-1)\omega_q+2d-5]rf'+(d-3)[(d-1)\omega_q+d-3](f-1)=0.\label{de1}
\end{eqnarray}
The general solution of the above equation has the form
\begin{eqnarray}
f=1-\frac{r_g}{r^{d-3}}+\frac{c_1}{r^{(d-1)\omega_q+d-3}},
\end{eqnarray}
where $r_g$ and $c_1$ are normalization factors. When $c_1=0$, the
function $f$ describes the usual $d$-dimensional Schwarzschild
black hole. Moreover, we also note that in the case $\omega_q=0$,
the second and the third term in $f$ have the same order of $r$.

The energy density $\rho_q$ for quintessence can be described by
\begin{eqnarray}
\rho_q=\frac{c_1\omega_q(d-1)(d-2)}{4r^{(d-1)(\omega_q+1)}}.
\end{eqnarray}
which should be positive. Since $\omega_q\leq 0$, it requires the
normalization constant $c_1$ for quintessence to be negative. If
we take $r_g=2M$ and $c=-c_1$, the metric of the $d$-dimensional
spherically symmetric black hole surrounded by quintessence reads
\begin{eqnarray}
ds^2=\bigg[1-\frac{2M}{r^{d-3}}-\frac{c}{r^{(d-1)\omega_q+d-3}}\bigg]dt^2
-\bigg[1-\frac{2M}{r^{d-3}}-\frac{c}{r^{(d-1)\omega_q+d-3}}\bigg]^{-1}dr^2
-r^2d\Omega_{d-2}.\label{metric1}
\end{eqnarray}
This spacetime depends not only on the dimension $d$, but also on
the state parameter $\omega_q$ of quintessence. When $d=4$, our
result reduces to that obtained in \cite{18}. In the limit
$\omega_q=-1$, the metric (\ref{metric1}) becomes
\begin{eqnarray}
ds^2=\bigg[1-\frac{2M}{r^{d-3}}-cr^2\bigg]dt^2
-\bigg[1-\frac{2M}{r^{d-3}}-cr^2\bigg]^{-1}dr^2 -r^2d\Omega_{d-2},
\end{eqnarray}
which reduces to the $d$-dimensional Schwarzschild de Sitter black
hole. We also note that the metric (\ref{metric1}) can reduce to
the $d$-dimensional Reissner-Nordstr\"{o}m black hole if we take
\begin{eqnarray}
\omega_q=\frac{d-3}{d-1}.
\end{eqnarray}
This implies that the state parameter $\omega_q$ of the
electromagnetic field is a function of the dimension $d$ of the
spacetime, so that we might fix the number of extra dimensions of
the spacetime by measuring the relation between the pressure $p_q$
and the energy density $\rho_q$.

\section{Thermodynamics of the $d$-dimensional static spherically-symmetric black hole surrounded by
quintessence}

We now study the thermodynamical property at the black hole event
horizon in the background (\ref{metric1}).

We write the mass $E$ of a $d$-dimensional black hole as a product
\begin{eqnarray}
 E=F(d)M,
\end{eqnarray}
where $F(d)$ is a function of dimension $d$. We will see that the
first law of thermodynamics at the black hole event horizon does
not depend on this function. The entropy $S$, mass $E$ and Hawking
temperature $T$ of the black hole (\ref{metric1}) can be described
by
\begin{eqnarray}
S&=&\frac{A_h}{4}=\frac{(d-1)\pi^{(d-1)/2}}{4\Gamma
[\frac{d+1}{2}]}r^{d-2}_h=\frac{r^{d-2}_h}{G(d)},\\
E&=&\frac{F(d)}{2}[G(d)S]^{\frac{d-3}{d-2}}-\frac{F(d)c}{2}[G(d)S]^{-\frac{\omega_q(d-1)}{d-2}},
\\
T&=&\frac{F(d)G(d)}{2(d-2)}\bigg[(d-3)[G(d)S]^{\frac{-1}{d-2}}+\omega_qc
(d-1)[G(d)S]^{-\frac{\omega_q(d-1)+d-2}{d-2}}\bigg],
\end{eqnarray}
respectively. As did in \cite{19}, we treat the constant $c$ as a
variable, and have the generalized force
\begin{eqnarray}
 \Theta_c=\bigg(\frac{\partial E}{\partial
 c}\bigg)_{S}=-\frac{F(d)}{2}[G(d)S]^{-\frac{\omega_q(d-1)}{d-2}}.
\end{eqnarray}
We find that the first law takes the form
\begin{eqnarray}
 \frac{d-3}{d-2}E=TS+\frac{\omega_q(d-1)+d-3}{d-2}\Theta_c
 \;c.\label{MFL1}
\end{eqnarray}
It is clear that the mass depends on the state parameter
$\omega_q$ of quintessence. In the limit $\omega_q\rightarrow -1$,
the second term in the right-side of equation (\ref{MFL1}) becomes
$-\frac{2}{d-2}\Theta_c \;c$. Setting $c=\frac{1}{l^2}$, where
$l^2$ is defined as the cosmological constant $\Lambda$ through
$l^2=\frac{(d-1)(d-2)}{2\Lambda}$, we have
$\Theta_l=-\frac{2}{l^3}\Theta_c$ and $-\frac{2}{d-2}\Theta_c
\;c=\frac{1}{d-2}\Theta_l\;l$. Then the equation (\ref{MFL1})
reduces to the first law in the $d$-dimensional de Sitter black
hole spacetimes \cite{19}. In the case $\omega_q=\frac{d-3}{d-1}$,
setting $c=-Q^2$, we can obtain $\Theta_Q=-2Q\Theta_c$, and
$\frac{2(d-3)}{d-2}\Theta_c \;c=\frac{d-3}{d-2}\Theta_Q \;Q$. The
first law returns to that in the $d$-dimensional
Reissner-Nordstr\"{o}m black hole\cite{20}.

From equations (26), (27) and (28), it is easy to obtain that
\begin{eqnarray}
T=\bigg(\frac{\partial E}{\partial S}\bigg)_c.
\end{eqnarray}
Combining it with equation (29), we have
\begin{eqnarray}
dE=\bigg(\frac{\partial E}{\partial S}\bigg)_c
dS+\bigg(\frac{\partial E}{\partial c}\bigg)_Sd\;c= T dS+\Theta_c
 d\;c,\label{MFL2}
\end{eqnarray}
which is the differential form of the first law of thermodynamics in
the background of (22). Obviously, in the case of a $d$-dimensional
static black hole surrounded by spherically-symmetric quintessence,
the differential form of the first law does not depend on the state
parameter $\omega_q$.

\section{Greybody factor for a $d$-dimensional static spherically-symmetric black hole surrounded by
quintessence}

In this section, we study the greybody factors for the emission of
scalar field in the low energy limit on the brane and into the
bulk from the $d$-dimensional static spherically-symmetric black
hole surrounded by quintessence (\ref{metric1}). The greybody
factors of the scalar field in the $d$-dimensional Schwarzschild
and Schwarzschild-dS blackholes have been investigated
\cite{TJR,p22}.

The equation of motion for a massless scalar particle propagating in
the curved spacetime is described by
\begin{eqnarray}
\frac{1}{\sqrt{-g}}\partial_{\mu}(\sqrt{-g}g^{\mu\nu}\partial_{\nu})
\Phi(t,r,\Omega)=0,\label{WE}
\end{eqnarray}
where $\Phi(t,r,\Omega)$ denotes the scalar field. Separating
$\Phi(t,r,\Omega)=e^{-i\omega t}\Psi_{bulk}(r)Y_{lm}(\Omega)$, we
can obtain the radial equation for the scalar field propagating
into the bulk
\begin{eqnarray}
\frac{1}{r^{d-2}}\frac{d}{dr}\bigg[r^{d-2}f\frac{d
\Psi_{bulk}(r)}{dr}\bigg]
+\bigg[\frac{\omega^2}{f}-\frac{l(l+d-3)}{r^2}\bigg]\Psi_{bulk}(r)=0,\label{radial}
\end{eqnarray}
with $f=1-\frac{2M}{r^{d-3}}-\frac{c}{r^{(d-1)\omega_q+d-3}}$.
Similarly, we can also obtain the radial equation for scalar field
propagating on the brane
\begin{eqnarray}
\frac{1}{r^{2}}\frac{d}{dr}\bigg[r^{2}f\frac{d
\Psi_{brane}(r)}{dr}\bigg]
+\bigg[\frac{\omega^2}{f}-\frac{l(l+1)}{r^2}\bigg]\Psi_{brane}(r)=0.\label{radia2}
\end{eqnarray}
Adopting the tortoise coordinate $x=\int{\frac{dr}{f}}$, radial
equations (\ref{radial}) and (\ref{radia2}) can be further written
as
\begin{eqnarray}
\bigg[\frac{d^2}{dx^2}+\omega^2-V_{bulk}(r)\bigg]\bigg[r^{\frac{d-2}{2}}\Psi_{bulk}(r)\bigg]=0,\label{radia3}
\end{eqnarray}
and
\begin{eqnarray}
\bigg[\frac{d^2}{dx^2}+\omega^2-V_{brane}(r)\bigg]\bigg[r\Psi_{brane}(r)\bigg]=0,\label{radia4}
\end{eqnarray}
with the effective potentials
\begin{eqnarray}
V_{bulk}(r)=f\bigg[\frac{(d-2)(d-4)f}{4r^2}+\frac{(d-2)f'}{2r}+\frac{l(l+d-3)}{r^2}\bigg],\label{Vbu}
\end{eqnarray}
and
\begin{eqnarray}
V_{brane}(r)=f\bigg[\frac{f'}{r}+\frac{l(l+1)}{r^2}\bigg].\label{Vbr}
\end{eqnarray}

Here we only consider the greybody factor for the mode $l=0$ which
dominates in the low-energy regime $\omega\ll T_H$ and $\omega
R_H\ll 1$.

From the expression of $f$, we find that the spacetime
(\ref{metric1}) is asymptotically flat if $0>\omega_q>-(d-3)/(d-1)$
and is asymptotically dS-like when $-1\leq\omega_q<-(d-3)/(d-1)$. As
in \cite{TJR}, we write $f=f_a(r)+f_h(r)$. The function $f_a(r)$ is
the asymptotic part of $f$ and the function $f_h(r)$ contains
physics which is specific for the black hole. We can define the
asymptotic region to be $f_a(r)\gg f_h(r)$.

Near the black hole horizon $r\sim R_H$, taking into account the
ingoing boundary condition, we obtain the solution of the radial
equations (\ref{radial}) and (\ref{radia2}) in the same form
\begin{eqnarray}
\Psi(r)_{RH}=A_Ie^{i\omega x}. \label{p1}
\end{eqnarray}
Near the black hole horizon $x\sim\frac{1}{2\kappa_H}
\log{\frac{r-R_H}{R_H}}$, the solution (\ref{p1}) can be written as
\begin{eqnarray}
\Psi(r)_{RH}=A_I
\bigg[1+\frac{i\omega}{2\kappa_H}\log{\frac{r-R_H}{R_H}}\bigg].\label{p11}
\end{eqnarray}

In the intermediate region where the effective potentials
$V_{bulk}(r)\gg \omega^2$ and $V_{brane}(r)\gg \omega^2$, the radial
equations (\ref{radial}) and (\ref{radia2}) can be reduced to
\begin{eqnarray}
\frac{1}{r^{d-2}}\frac{d}{dr}\bigg[r^{d-2}f\frac{d
\Psi_{bulk}(r)}{dr}\bigg]=0, \label{ri1}
\end{eqnarray}
and
\begin{eqnarray}
\frac{1}{r^{2}}\frac{d}{dr}\bigg[r^{2}f\frac{d
\Psi_{brane}(r)}{dr}\bigg]=0,\label{ri2}
\end{eqnarray}
respectively. The general solutions for equation (\ref{ri1}) and
(\ref{ri2}) can be got
\begin{eqnarray}
\Psi_{bulk}(r)=A_{II}+B_{II}G(r),\;\;\;\;\;\;\;\text{and}\;\;\;\;\;\;
\Psi_{brane}(r)=A_{II}+B'_{II}G'(r),\label{rs2}
\end{eqnarray}
where
\begin{eqnarray}
G(r)=\int^r_{\infty}\frac{dr}{r^{d-2}f},\;\;\;\;\;\;\;\text{and}\;\;\;\;\;\;G'(r)=\int^r_{\infty}\frac{dr}{r^{2}f}.
\end{eqnarray}
For $r\sim R_H$,  we have
\begin{eqnarray}
G(r)=\frac{1}{2R^{d-2}_H\kappa_H}\log{(r-R_H)},\;\;\;\;\;\;\;\text{and}\;\;\;\;\;\;G'(r)=\frac{1}{2R^{2}_H\kappa_H}\log{(r-R_H)}.
\end{eqnarray}
Matching this solution to the wave-function (\ref{p11}) of the
near black hole horizon region, we obtain
\begin{eqnarray}
A_{II}=A_I,\;\;\;\;\;\;\;\;\;\; B_{II}=i\omega
R^{d-2}_HA_I,\;\;\;\;\;\;\;\;\;\; B'_{II}=i\omega R^{2}_HA_I.
\end{eqnarray}
The asymptotic expressions of wave functions (\ref{rs2}) in the
limit $r\gg R_H$ with $V(r)\gg \omega^2$ read
\begin{eqnarray}
\Psi_{bulk}(r)=A_{I}\bigg(1+i\omega R^{d-2}_H
\int^r_{\infty}\frac{dr}{r^{d-2}f_a(r)}\bigg),\;\;\; \;\;\;\;
\Psi_{brane}(r)=A_{I}\bigg(1+i\omega R^{2}_H
\int^r_{\infty}\frac{dr}{r^{2}f_a(r)}\bigg). \label{rs22}
\end{eqnarray}
We shall use these expressions to match the general solution for
the scalar wave equation in the asymptotic region.

Until now we just concentrate on the black hole, in the following
we will do the matching in the asymptotic region for the case of
asymptotically flat spacetime and asymptotically dS spacetime,
respectively.

Let us first consider the asymptotically flat spacetime where
$0>\omega_q>-(d-3)/(d-1)$. In this case, we take $f_a(r)=1$.

The general solutions of the wave equation (\ref{radial}) and
(\ref{radia2}) in asymptotically flat spacetime are given by
\begin{eqnarray}
\Psi_{bulk}(r)=\rho^{\frac{3-d}{2}}\bigg[C_1H^{(1)}_{\frac{d-3}{2}}(\rho)+C_2H^{(2)}_{\frac{d-3}{2}}(\rho)\bigg],
\;\;\; \;\;\;\;
\Psi_{brane}(r)=\rho^{-\frac{1}{2}}\bigg[C_1H^{(1)}_{\frac{1}{2}}(\rho)+C_2H^{(2)}_{\frac{1}{2}}(\rho)\bigg],
\end{eqnarray}
where $\rho=r \omega$,
$H^{(1)}_{\nu}(\rho)=J_{\nu}(\rho)+iN_{\nu}(\rho)$ and
$H^{(2)}_{\nu}(\rho)=J_{\nu}(\rho)-iN_{\nu}(\rho)$ are the Hankel
functions defined by the Bessel functions $J_{\nu}(\rho)$ and
$N_{\nu}(\rho)$. In the limit $\rho\ll 1$, we have
\begin{eqnarray}
\Psi_{bulk}(r)\sim
\frac{C_1+C_2}{\Gamma(\frac{d-1}{2})2^{\frac{d-3}{2}}}-i(C_1-C_2)\frac{\Gamma(\frac{d-3}{2})2^{\frac{d-3}{2}}}{\pi
\rho^{\frac{d-3}{2}}},\;\;\; \;\;\;\; \Psi_{brane}(r)\sim
\frac{\sqrt{2}(C_1+C_2)}{\sqrt{\pi}}-\frac{i\sqrt{2}(C_1-C_2)}{\sqrt{\pi}
\rho^{\frac{d-3}{2}}}.\label{rs31}
\end{eqnarray}
\begin{table}[h]
\begin{center}
\begin{tabular}[b]{c|c|c|c|c|c|c}
 \hline \hline
 &\multicolumn{2}{c}{$R_H$} &&\multicolumn{2}{c}{$R^{d-2}_H$ }
 \\
\hline &$\omega_q=-0.1$&
 $\omega_q=-0.2$& $\omega_q=-0.3$&
 $\omega_q=-0.1$& $\omega_q=-0.2$&
 $\omega_q=-0.3$
\\ \hline & & & & & &
\\
$d=4$& 2.0122
 &2.0151&2.0188&4.0481
 &4.0602& 4.0756
\\& & & & & &
\\
$d=5$&1.4183
 &1.4189&1.4200&2.8530
 &2.8566&2.8633
 \\& & && & & \\
$d=6$&1.2623
 &1.2626&1.2629&2.5389& 2.5413
 &2.5438
\\& & && & &
\\
$d=7$&1.1909
 &1.1910&1.1912&2.3954
 &2.3964&2.3984
\\ \hline\hline
\end{tabular}
\caption{The changes of $R_H$ and $R^{d-2}_H$ with different state
parameter $\omega_q$ and dimension numbers $d$ in the asymptotically
flat case. Here $M=1$ and $c=0.01$.}
\end{center}
\end{table}

Matching the wave-function (\ref{rs31}) in the asymptotic region
to that in the intermediate region, we get the relationship
between the coefficients $C_1$ and $C_2$
\begin{eqnarray}
C_1+C_2=\Gamma(\frac{d-1}{2})2^{\frac{d-3}{2}}A_I,\;\;\;\;\;\;C_1-C_2=\frac{\pi
\omega^{d-2}R^{d-2}_H}{(d-3)\Gamma(\frac{d-3}{2})2^{\frac{d-3}{2}}}A_I,
\end{eqnarray}
in the bulk and
\begin{eqnarray}
C_1+C_2=\sqrt{\frac{\pi}{2}}A_I,\;\;\;\;\;\;C_1-C_2=\sqrt{\frac{\pi}{2}}\omega^2R^2_HA_I,
\end{eqnarray}
on the brane. From the definition of greybody factor in the low
energy limit $\omega R_H\ll 1$,
\begin{eqnarray}
\gamma (\omega)=1-\frac{|C_2|^2}{|C_1|^2}\simeq
4\frac{C_1-C_2}{C_1+C_2},
\end{eqnarray}
we obtain the greybody factor
\begin{eqnarray}
\gamma (\omega)=\frac{4\pi
\omega^{d-2}R^{d-2}_H}{2^{d-2}\Gamma(\frac{d-1}{2})^2},\label{GFb1}
\end{eqnarray}
in the bulk and
\begin{eqnarray}
\gamma(\omega)=4\omega^{2}R^{2}_H,\label{GFb2}
\end{eqnarray}
on the brane. Obviously, the greybody factors in the bulk and on the
brane depend on the black hole horizon radius. The changes of $R_H$
and $R^{d-2}_H$ with the state parameter $\omega_q$ and dimension
$d$ are listed in table (I) and the factor
$2^{d-2}\Gamma(\frac{d-1}{2})^2$ in (\ref{GFb1}) is only a
monotonically increased function of the dimension numbers $d$, thus
the greybody factors (\ref{GFb1}) and (\ref{GFb2}) increase with the
increase of the absolute value of $\omega_q$ and decrease with the
increase of $d$ .

In the $d$ dimensional black hole spacetime, the luminosity of the
black hole Hawking radiation for the mode $l=0$ in the bulk and on
the brane is given by
\begin{eqnarray}
L_{bulk}&=&\int^{\infty}_0\frac{d\omega\;
\omega^{d-1}R^{d-2}_H}{2^{d-3}\Gamma(\frac{d-1}{2})^2}\frac{
1}{e^{\;\omega/T_{H}}-1},\\
L_{brane}&=&\int^{\infty}_0\frac{d\omega}{2\pi}
\frac{4\omega^3R^{2}_H}{e^{\;\omega/T_{H}}-1}.
\end{eqnarray}
The integral expressions above are just for the sake of
completeness by writing the integral range from $0$ to infinity.
However, as our analysis has focused only in the low-energy regime
of the spectrum, an upper cutoff will be imposed on the energy
parameter such that the low-energy conditions $\omega \ll T_H$ and
$\omega R \ll 1$ are satisfied. The values derived for the
luminosities of the black hole on the brane and in the bulk will
therefore be based on the lower part of the spectrum and
modifications may appear when the high-energy part of the spectrum
is included in the calculation.
\begin{table}[h]
\begin{center}
\begin{tabular}[b]{ccccccc}
 \hline \hline
 \;\;\;\; \;\;\;\; & \;$\omega_q=-0.1$\;\;\;\;
& \;\;\;\;$\omega_q=-0.2$ \;\;\;\;& \;\;$\omega_q=-0.3$  &
\;\;$\omega_q=-0.4$& \;\;$\omega_q=-0.5$& \;\;$\omega_q=-0.6$ \\
\hline
\\
$d=4$&\;\;\;\;\;0.03947\;\;\;\;\;& \;\;\;\; 0.03931\;\;\;\;\;
 &\;\;\;\;\;0.03909\;\;\;\;& \;\;\;\;\;\;\;\;\;\;
 &\;\;\;\;\;\;\;\;\;&\;\;\;\;\;\;\;\;\;
\\
\\
$d=5$&\;\;\;\;\;0.11208\;\;\;\;\;& \;\;\;\;0.11187\;\;\;\;\;
 &\;\;\;\;\;0.11161\;\;\;\;& \;\;\;\;\;0.11127\;\;\;\;\;
 &\;\;\;\;0.11086\;\;\;\;\;
 \\
 \\
$d=6$&\;\;\;\;\;0.18895\;\;\;\;\;& \;\;\;\;\;0.18869\;\;\;\;\;&
\;\;\;\; 0.18837\;\;\;\;\;
 & \;\;\;\;0.18798\;\;\;\;\;&\;\;\;\;0.18752\;\;\;\;\;&\;\;\;\;0.18696\;\;\;\;\;
 \\
\\
$d=7$&\;\;\;\;\;0.26707\;\;\;\;\;& \;\;\;\; 0.26676 \;\;\;\;\;
 &\;\;\;\;\;0.26639\;\;\;\;& \;\;\;\;\;0.26595\;\;\;\;\;&\;\;\;\;0.26542\;\;\;\;\;&\;\;\;\;0.26480\;\;\;\;\;
\\ \hline\hline
\end{tabular}
\caption{The change of $T_H$ with different state parameter
$\omega_q$ and dimension numbers $d$ in the asymptotically flat
case. Here $M=1$ and $c=0.01$.}
\end{center}
\end{table}
The Hawking temperature $T_H$ of the black hole in the
asymptotically flat spacetime is listed in table (II). It is shown
that $T_H$ increases with of the dimension number $d$ and decreases
with the increase of the absolute value of $\omega_q$. Table (III)
tells us that both the luminosity of Hawking radiation in the bulk
and on the brane decrease with the increase of the absolute
$\omega_q$ and increase with the dimension number $d$. We observe
that Hawking radiation dominates on the brane and the ratio
$L_{brane}/L_{bulk}$ increases with the magnitude of $\omega_q$ and
dimension $d$.
\begin{table}[h]
\begin{center}
\begin{tabular}[b]{c|c|c|c|c|c|c|c|c|c}
 \hline \hline
 &\multicolumn{2}{c}{$L_{bulk}\;\;(10^{-5})$} &&\multicolumn{2}{c}{$L_{brane} \;\;(10^{-5})$ }
 &&\multicolumn{2}{c}{$L_{brane}/L_{bulk}$ }
 \\
\hline &
 $\omega_q=-0.1$& $\omega_q=-0.2$&
 $\omega_q=-0.3$&
 $\omega_q=-0.1$& $\omega_q=-0.2$&
 $\omega_q=-0.3$&
 $\omega_q=-0.1$& $\omega_q=-0.2$&
 $\omega_q=-0.3$
\\ \hline & & & & & & & & &
\\
$d=4$& $4.06400$ &$4.0087$
 & $3.9330$& $--$ &$--$
 & $--$
 & $--$& $--$&$--$
\\& & && & & & & &
\\
$d=5$&$31.405$ &$31.145$
 & $30.818$& $131.27$&$130.39$
 & $129.26$
 & $4.180$& 4.186&4.194
 \\& & && & & & & &\\
$d=6$&$99.774$
 & $99.036$
 &$98.135$
 &$839.67$&$835.39$& $830.17$&
  $8.416$&$8.435$&8.460
\\& & && & & & & &
\\
$d=7$&$263.32$
 &$261.38$
 &$259.06$
 &$2982.9$&$2969.9$& $2954.4$&
  $11.33$&$11.36$&11.40
\\ \hline\hline
\end{tabular}
\caption{The changes of $L_{bulk}$, $L_{brane}$ and
$L_{brane}/L_{bulk}$ with different state parameter $\omega_q$ and
dimension numbers $d$ in the asymptotically flat case. Here $M=1$
and $c=0.01$.}
\end{center}
\end{table}

Now we start to consider asymptotically dS like spacetime with
$-1\leq\omega_q<-(d-3)/(d-1)$. The function $f_a(r)$ is now given by
\begin{eqnarray}
 f_a(r)=1-\frac{c}{r^{\omega_q(d-1)+d-3}},
\end{eqnarray}
The metric (\ref{metric1}) now has a cosmological-like horizon
located at $r=r_c=c^{1/[\omega_q(d-1)+d-3]}$. Assuming that $r_c\gg
R_H$, we have that $f_h(r)\ll f_a(r)$ for $r\gg R_H$ and the
$f_h(r)$ contribution to $f(r)$ is negligible. This allows us to
define an intermediate region, $R_H\ll r\ll r_c$, in between the
near horizon and the asymptotic region. Thus, for $r\gg R_H$,
$r/r_c\ll 1$ and $r\omega\ll 1$, the wave functions (\ref{rs22})
have the form
\begin{eqnarray}
\Psi_{bulk}(r)=A_{I}\bigg(1-i\frac{\omega R^{d-2}_H}{(d-3)r^{d-3}}
\bigg),\;\;\; \;\;\;\; \Psi_{brane}(r)=A_{I}\bigg(1-i\frac{\omega
R^{2}_H}{r} \bigg). \label{rds3}
\end{eqnarray}

In the asymptotic region, $r\gg R_H$, we define the coordinate
\begin{eqnarray}
z=\bigg(\frac{r}{r_c}\bigg)^{-n},
\end{eqnarray}
with $n=\omega_q(d-1)+d-3$, which is negative because that in this
asymptotically dS-like spacetime $\omega_q<-(d-3)/(d-1)$. Then
radial equations (\ref{radial}) and (\ref{radia2}) can be
approximated as
\begin{eqnarray}
(1-z)z\frac{d^2
P_{bulk}}{dz^2}-\bigg[\frac{n+1}{n}-z\frac{2n+1}{n}\bigg]\frac{dP_{bulk}}{dz}
+\bigg\{\frac{\omega^2
r^2_c}{n^2(1-z)}-\frac{d-2}{4zn^2}\bigg[d-4-z(d-2n-4)\bigg]\bigg\}P_{bulk}=0,\label{bulkf}
\end{eqnarray}
and
\begin{eqnarray}
(1-z)z\frac{d^2
P_{brane}}{dz^2}-\bigg[\frac{n+1}{n}-z\frac{2n+1}{n}\bigg]\frac{dP_{brane}}{dz}
+\bigg[\frac{\omega^2
r^2_c}{n^2(1-z)}-\frac{1}{n}\bigg]P_{brane}=0,\label{branef}
\end{eqnarray}
respectively. Here $P_{bulk}=r^{(d-2)/2}\Psi_{bulk}(r)$ and
$P_{brane}=r\Psi_{brane}(r)$.

The general solution to the equation (\ref{bulkf}) is
\begin{eqnarray}
P_{bulk}&=&C_1z^{-\frac{d-2}{2n}}(1-z)^{\frac{i\omega
r_c}{n}}\;_2F_1\bigg[\frac{i\omega
r_c}{n},\frac{3-d+n}{n}+\frac{i\omega r_c}{n},
\frac{3-d+n}{n};z\bigg]\nonumber\\
&+& C_2z^{\frac{d-4}{2n}}(1-z)^{\frac{i\omega
r_c}{n}}\;_2F_1\bigg[1+\frac{i\omega
r_c}{n},\frac{d-3}{n}+\frac{i\omega r_c}{n},
\frac{d-3+n}{n};z\bigg], \label{bulkfs}
\end{eqnarray}
where $_2F_1[a,b,\tilde{c};z]$ is the standard hypergeometric
function. Since, $n<0$, for $z\rightarrow 0$, or $r/r_c\ll 1$, we
have
\begin{eqnarray}
\Psi_{bulk}(r)=C_1r^{\frac{2-d}{2}}_c+\frac{C_2r^{\frac{d-4}{2}}_c}{r^{d-3}}.
\end{eqnarray}
Matching this wave function to the behavior (\ref{rds3}) in the
intermediate region, we can fix the coefficients $C_1$ and $C_2$
\begin{eqnarray}
C_1=r^{\frac{d-2}{2}}_cA_I,\;\;\;\;\;\;\;\;C_2=-ir^{\frac{4-d}{2}}_c\frac{\omega
R^{d-2}_H}{(d-3)}A_I.
\end{eqnarray}
In order to find the behavior of the wave function for $z\rightarrow
1$, we change the argument of the hypergeometric function of the
solution (\ref{bulkfs}) from $z$ to $1-z$ and find that it has the
form
\begin{eqnarray}
P_{bulk}&=&C_1 b_{11}z^{-\frac{d-2}{2n}}(1-z)^{\frac{i\omega
r_c}{n}}\;_2F_1\bigg[\frac{i\omega r_c}{n},\frac{3-d+n+i\omega
r_c}{n},
1+\frac{i2\omega r_c}{n};1-z\bigg]\nonumber\\
&+& C_1b_{21}z^{-\frac{d-2}{2n}}(1-z)^{-\frac{i\omega
r_c}{n}}\;_2F_1\bigg[\frac{3-d+n-i\omega r_c}{n},-\frac{i\omega
r_c}{n}, 1-\frac{i2\omega r_c}{n};1-z\bigg]
\nonumber\\
&+& C_2b_{12}z^{\frac{d-4}{2n}}(1-z)^{\frac{i\omega
r_c}{n}}\;_2F_1\bigg[1+\frac{i\omega r_c}{n},\frac{d-3+i\omega
r_c}{n}, 1+\frac{i2\omega
r_c}{n};1-z\bigg]\nonumber\\
&+& C_2b_{22}z^{\frac{d-4}{2n}}(1-z)^{-\frac{i\omega
r_c}{n}}\;_2F_1\bigg[\frac{d-3-i\omega r_c}{n}, 1-\frac{i\omega
r_c}{n},1-\frac{i2\omega r_c}{n};1-z\bigg], \label{bulkfs1}
\end{eqnarray}
with
\begin{eqnarray}
b_{11}=\frac{\Gamma(\frac{3-d+n}{n})\Gamma(-\frac{i2\omega
r_c}{n})}{\Gamma(\frac{3-d+n-i\omega r_c}{n})\Gamma(-\frac{i\omega
r_c}{n})},\;\;\;\;\;\;\;
b_{12}=\frac{\Gamma(\frac{d-3+n}{n})\Gamma(-\frac{i2\omega
r_c}{n})}{\Gamma(\frac{d-3-i\omega r_c}{n})\Gamma(1-\frac{i\omega
r_c}{n})},
\end{eqnarray}
\begin{eqnarray}
b_{21}=\frac{\Gamma(\frac{3-d+n}{n})\Gamma(\frac{i2\omega
r_c}{n})}{\Gamma(\frac{i\omega r_c}{n})\Gamma(\frac{n+3-d+i\omega
r_c}{n})},\;\;\;\;\;\;\;
b_{22}=\frac{\Gamma(\frac{d-3+n}{n})\Gamma(\frac{i2\omega
r_c}{n})}{\Gamma(\frac{d-3+i\omega r_c}{n})\Gamma(1+\frac{i\omega
r_c}{n})}.
\end{eqnarray}
Thus, in the limit $z\rightarrow 1$, the wave function becomes
\begin{eqnarray}
\Psi_{bulk}(r)=\tilde{C_1}r^{\frac{2-d}{2}}_ce^{-\frac{i\omega r_c
\delta}{n}}e^{i\omega
x}+\tilde{C_2}r^{\frac{2-d}{2}}_ce^{\frac{i\omega r_c
\delta}{n}}e^{-i\omega x},
\end{eqnarray}
where
$\delta=-\frac{\Gamma(1-1/n)}{\Gamma(-1/n)}[\text{EulerGamma}+\text{PolyGamma}(0,-1/n)]$.
The relations between $\tilde{C_1}$, $\tilde{C_2}$ and $C_1$, $C_2$
can be expressed as
\begin{eqnarray}
\bigg(
\begin{array}{l}\tilde{C_1}\\
\tilde{C_2}
\end{array}\bigg)
=\bigg(
\begin{array}{l}b_{11}\;\;\;\;\; b_{12}\\
b_{21}\;\;\;\;\; b_{22}
\end{array}\bigg)\bigg(
\begin{array}{l}C_1\\
C_2
\end{array}\bigg).
\end{eqnarray}
Then the greybody factor is give by
\begin{eqnarray}
\gamma(\omega)_{bulk}&=&1-\frac{|\tilde{C_2}|^2}{|\tilde{C_1}|^2}=\bigg|\frac{b_{21}}{b_{11}}\bigg|^2
\bigg|1-\frac{b_{11}b_{22}-b_{12}b_{21}}{b_{11}b_{21}}\frac{C2}{C1}\bigg|^2\nonumber\\
&=&4h(\omega r_c)\bigg(\frac{R_H}{r_c}\bigg)^{d-2},
\end{eqnarray}
where the function $h(\omega r_c)$ is defined by
\begin{eqnarray}
h(\omega r_c)=\frac{1}{4|b_{11}|^2}.
\end{eqnarray}
Similarly, the greybody factor for the scalar emission on the
brane can be got
\begin{eqnarray}
\gamma(\omega)_{brane}=\frac{1}{|b'_{11}|^2}\bigg(\frac{R_H}{r_c}\bigg)^{2},
\end{eqnarray}
with
\begin{eqnarray}
b'_{11}=\frac{\Gamma(\frac{n-1}{n})\Gamma(-\frac{i2\omega
r_c}{n})}{\Gamma(\frac{n-1-i\omega r_c}{n})\Gamma(-\frac{i\omega
r_c}{n})}.
\end{eqnarray}
The greybody factors $\gamma(\omega)_{bulk}$ and
$\gamma(\omega)_{brane}$ depend on $\omega$ and the ratio
$R_H/r_c$. The changes of $R_H$ and $r_c$ in the case
$\omega_q(d-1)+d-3<0$ with $\omega_q$ and $d$ are listed in table
(IV). One can find that the ratio $R_H/r_c$ increases with the
increase of the absolute value of $\omega_q$ and decreases with
the increase of the dimension $d$. In the low energy limit $\omega
r_c<1$, we have the quantities $|b_{11}|^2\sim 1/4$ and
$|b'_{11}|^2\sim 1/4$, and then the greybody factors
$\gamma(\omega)_{bulk}$ and $\gamma(\omega)_{brane}$ also increase
with $\omega_q$ and decrease with $d$.
\begin{table}[h]
\begin{center}
\begin{tabular}[b]{c|c|c|c|c|c|c|c|c|c}
 \hline \hline
 &\multicolumn{2}{c}{$R_H$} &&\multicolumn{2}{c}{$r_c$ }&&\multicolumn{2}{c}{$R_H/r_c$ }
 \\
\hline &
 $\omega_q=-0.8$& $\omega_q=-0.9$&
 $\omega_q=-1$&
 $\omega_q=-0.8$& $\omega_q=-0.9$&
 $\omega_q=-1$&
 $\omega_q=-0.8$& $\omega_q=-0.9$&
 $\omega_q=-1$
\\ \hline & & & & & & & &&
\\
$d=4$&$2.0564$
 & $2.0714$& $2.0915$
 &$25.294$
 & $13.680$& $8.789$&0.0813&0.1514&0.2380
\\& & && & & & &&
\\
$d=5$
 & $1.4252$
 &$1.4269$
 &$1.4289$& $46.380$&
  $17.712$&$9.897$&0.0307&0.0806&0.1444
 \\& & && & & & && \\
$d=6$
 & $1.2653$
 &$1.2660$
 &$1.2667$& $100.00$&
  $21.542$&$9.990$&0.0127&0.0588&0.1268
\\& & && & & & &&
\\
$d=7$
 &$1.1927$
 &$1.1930$
 &$1.1935$& $316.23$&
  $26.827$&$9.999$&0.0038&0.0445&0.1194
\\ \hline\hline
\end{tabular}
\caption{The changes of $R_H$, $r_c$ and $R_H/r_c$with different
state parameter $\omega_q$ and dimension numbers $d$ in the
asymptotically dS-like case. Here $M=1$ and $c=0.01$.}
\end{center}
\end{table}

The luminosity of Hawking radiation for the mode $l=0$ in the bulk
and on the brane can be given by
\begin{eqnarray}
L_{bulk}&=&\int^{\infty}_0\frac{d\omega\;
\omega\;\gamma(\omega)_{bulk}}{e^{\;\omega/T_{H}}-1},\\
L_{brane}&=&\int^{\infty}_0\frac{d\omega}{2\pi} \frac{\omega\;
\gamma(\omega)_{brane}}{e^{\;\omega/T_{H}}-1}.
\end{eqnarray}
As did in (56)(57), here we also focus only in the low-energy
regime of the spectrum and impose an upper cutoff on the energy
parameter such that the low-energy conditions $\omega \ll T_H$ and
$\omega R \ll 1$ are satisfied.  The values obtained for the
luminosities of the black hole on the brane and in the bulk are
based on the lower part of the spectrum. When the high energy part
spectrum is included, the results may significantly change.

\begin{table}[h]
\begin{center}
\begin{tabular}[b]{cccccccc}
 \hline \hline
 \;\;\;\; \;\;\;\; & \;$\omega_q=-0.4$\;\;\;\;
& \;\;\;\;$\omega_q=-0.5$ \;\;\;\;& \;\;$\omega_q=-0.6$  &
\;\;$\omega_q=-0.7$& \;\;$\omega_q=-0.8$& \;\;$\omega_q=-0.9$& \;\;$\omega_q=-1.0$ \\
\hline
\\
$d=4$&\;\;\;\;\;0.03879\;\;\;\;\;& \;\;\;\; 0.03838\;\;\;\;\;
 &\;\;\;\;\;0.03784\;\;\;\;& \;\;\;\;0.03712\;\;\;\;\;\;
 &\;\;\;\;0.03615\;\;\;\;\;&\;\;\;\;0.03484\;\;\;\;\;&\;\;\;\;0.03306\;\;\;\;\;
\\
\\
$d=5$&\;\;\;\;\;\;\;\;\;\;& \;\;\;\;\;\;\;\;\;
 &\;\;\;\;\;0.11034\;\;\;\;& \;\;\;\;\;0.10971\;\;\;\;\;
 &\;\;\;\;0.10894\;\;\;\;\;&\;\;\;\;0.10800\;\;\;\;\;&\;\;\;\;0.10684\;\;\;\;\;
 \\
 \\
$d=6$&\;\;\;\;\;\;\;\;\;\;& \;\;\;\;\;\;\;\;\;\;& \;\;\;\;
\;\;\;\;\;
 & \;\;\;\;0.18629\;\;\;\;\;&\;\;\;\;0.18550\;\;\;\;\;&\;\;\;\;0.18455\;\;\;\;\;&\;\;\;\;0.18342\;\;\;\;\;
 \\
\\
$d=7$&\;\;\;\;\;\;\;\;\;\;& \;\;\;\; \;\;\;\;\;
 &\;\;\;\;\;\;\;\;\;& \;\;\;\;\;0.26407\;\;\;\;\;&\;\;\;\;0.26320\;\;\;\;\;&\;\;\;\;0.26219\;\;\;\;\;
 &\;\;\;\;0.26101\;\;\;\;\;
\\ \hline\hline
\end{tabular}
\caption{The change of $T_H$ with different state parameter
$\omega_q$ and dimensional numbers $d$ in the asymptotically dS-like
csse. Here $M=1$ and $c=0.01$.}
\end{center}
\end{table}
\begin{table}[h]
\begin{center}
\begin{tabular}[b]{c|c|c|c|c|c|c|c|c|c}
 \hline \hline
 &\multicolumn{2}{c}{$L_{bulk}\;\;(10^{-5})$} &&\multicolumn{2}{c}{$L_{brane} \;\;(10^{-5})$ }
 &&\multicolumn{2}{c}{$L_{brane}/L_{bulk}$ }
 \\
\hline &
 $\omega_q=-0.8$& $\omega_q=-0.9$&
 $\omega_q=-1.0$&
 $\omega_q=-0.8$& $\omega_q=-0.9$&
 $\omega_q=-1.0$&
 $\omega_q=-0.8$& $\omega_q=-0.9$&
 $\omega_q=-1.0$
\\ \hline & & & & & & & & &
\\
$d=4$& $0.9045$ &$2.9143$
 & $6.4797$&$--$ &$--$
 & $--$
 & $--$& $--$&$--$
\\& & && & & & & &
\\
$d=5$&$0.0361$ &$0.6386$
 & $3.5966$& $1.1734$&$7.9265$
 & $24.912$
 & $32.544$& 12.413&6.9267
 \\& & && & & & & &\\
$d=6$&$9.2\times 10^{-5}$
 & $0.0425$
 &$0.9108$
 &$0.5769$&$12.318$& $56.647$&
  $6246.3$&$289.54$&62.195
\\& & && & & & & &
\\
$d=7$&$5.5\times 10^{-9}$
 &$0.0013$
 &$0.1728$
 &$0.1032$&$14.238$& $101.64$&
  $1.864\times 10^{7}$&$11369.5$&588.06
\\ \hline\hline
\end{tabular}
\caption{The changes of $L_{bulk}$, $L_{brane}$ and
$L_{brane}/L_{bulk}$ with different state parameter $\omega_q$ and
dimensional numbers $d$ in the asymptotically dS-like case. Here
$M=1$ and $c=0.01$.}
\end{center}
\end{table}
Although the black hole Hawking temperature $T_H$ (which is listed
in table V) decreases with the increase of the absolute value of
$\omega_q$, table (VI) tells us that the luminosity of the black
hole Hawking radiation increases with the increase of the magnitude
of $\omega_q$. This is different from that in the asymptotically
flat case with $\omega_q(d-1)+d-3>0$. Moreover comparing with the
asymptotically flat case, although the ratio $L_{brane}/L_{bulk}$
tells us that the black hole Hawking radiation still dominates on
the brane in the dS like spacetime, its dependence on $|\omega_q|$
is different in the dS like situation from that in the
asymptotically flat spacetime. These differences can be understood
from the behavior of the ratio $R_H/r_c \ll 1$, which increases with
$|\omega_q|$. This means that when $|\omega_q|$ becomes bigger, the
black hole horizon and the cosmological horizon will come closer, so
that Hawking radiation on the black hole event horizon will be
enhanced by the contribution from Hawking radiation from the
cosmological horizon.

\section{conclusions and discussions}

In this paper, we obtain an exact solution of Einstein equations
for the static spherically-symmetric quintessential matter
surrounding a black hole in $d$-dimensional spacetimes. For
different state parameters $\omega_q$ of quintessence, our
solution can lead to different limits, such as the Schwarzschild,
Reissner-Nordstr\"{o}m and de Sitter black holes in
$d$-dimensions.  We study the thermodynamics in this d-dimensional
black hole spacetime and find that the first law is universal for
arbitrary state parameter $\omega_q$ of the quintessence.

We investigate the greybody factors and Hawking radiations of a
scalar field in the bulk and on the brane,  in the low-energy
regime, in this $d$-dimensional black hole surrounded by
quintessence. We observe that Hawking radiation dominates on the
brane. For the case $0>\omega_q>-(d-3)/(d-1)$, the black hole is
asymptotically flat, the luminosity of Hawking radiation both in
the bulk and on the brane decreases with the increase of
$|\omega_q|$. But for the case $-(d-3)/(d-1)>\omega_q>-1$, the
black hole is in the asymptotically dS spacetime, Hawking
radiation increases with the magnitude of $|\omega_q|$. The
difference can be attributed to the different asymptotic
structures of the spacetimes. In the asymptotic dS spacetime,
besides the black hole event horizon, there also exists the
cosmological horizon. When the absolute value of $\omega_q$
becomes bigger, these two horizons come closer. The contribution
of Hawking radiation from the cosmological horizon enhances the
Hawking radiation near the black hole event horizon.

\begin{acknowledgments}

This work was partially supported by NNSF of China, Shanghai
Education Commission and Shanghai Science and Technology Commission.
R. K. Su's work was partially supported by the National Basic
Research Project of China. S. B. Chen's work was partially supported
by the China Postdoctoral Science Foundation under Grant No.
20070410685, the Scientific Research Fund of Hunan Provincial
Education Department Grant No. 07B043, the National Basic Research
Program of China under Grant No. 2003CB716300 and the construct
program of key disciplines in Hunan Province.
\end{acknowledgments}

\end{document}